\documentclass[11pt]{article}

\usepackage[margin=1in]{geometry}
\usepackage[T1]{fontenc}
\usepackage[utf8]{inputenc}
\usepackage{lmodern}
\usepackage{amsmath,amssymb}
\usepackage{graphicx}
\usepackage{authblk}
\usepackage{float}
\usepackage{cite}
\usepackage{microtype}
\usepackage[hidelinks]{hyperref}

\title{Machine Learning to Foundation Models: Artificial Intelligence for Nanophotonic Modeling and Scientific Discovery}

\author[1]{Chaobin Yang}
\author[2]{Xueqing Liu\thanks{Corresponding author: \href{mailto:lxqlxq21@gmail.com}{lxqlxq21@gmail.com}}}
\author[3]{Yiqun Fu}
\author[4]{Fengbo Zhou}
\author[1]{Krzysztof Kempa}
\author[1]{Stefano Anzellotti}
\author[1]{Michael J. Naughton}

\affil[1]{Boston College, Chestnut Hill, MA 02467, USA}
\affil[2]{Columbia University, New York, NY 10027, USA}
\affil[3]{Carnegie Mellon University, Pittsburgh, PA 15213, USA}
\affil[4]{Washington University School of Medicine, St. Louis, MO 63110, USA}

\date{}

\begin{document}
\maketitle

\begin{abstract}
Artificial intelligence (AI) is increasingly used to model, design, and
study nanophotonic systems. This review traces the development of the
field from classical machine learning and deep learning to generative
models, transfer learning, transformers, and emerging foundation models.
It first introduces major nanophotonic platforms, including
nanoparticles, nanoholes, metasurfaces, photonic crystals, multilayer
thin films, and integrated photonic devices, together with their main
forward and inverse problems. It then reviews data-driven methods for
predicting optical spectra and fields, generating structures from target
responses, improving designs through optimization, and accounting for
fabrication constraints. Generative models are discussed as a way to
produce multiple valid solutions to nonunique inverse problems, while
transfer learning, few-shot learning, and physics-aware training help
reduce data requirements and improve generalization. Recent
domain-specific foundation models show that different optical structures
and responses can be handled within shared representations, but current
systems remain limited in scope and physical grounding. Future progress
will depend on multimodal models that connect geometry, materials,
spectra, electromagnetic (EM) fields, fabrication data, experiments, and
scientific literature with reliable simulation and validation tools.
Current foundation models remain domain-specific, and their extension to
broader nanophotonic tasks will require stronger physical grounding and
validation.
\end{abstract}


\section{Nanophotonic Modeling and Design Problems}

Nanophotonics encompasses diverse subwavelength structures that
manipulate light through scattering, interference, resonant confinement,
and guided-wave propagation. The optical response of spherical
nanoparticles was placed on a rigorous EM foundation by Mie's exact
solution for light scattering from homogeneous spheres\cite{ref1}, later
extended by Aden and Kerker to concentric core--shell particles\cite{ref2}.
Subsequently, experimental observations of strong electric- and
magnetic-dipole resonances in high-index silicon nanoparticles
established dielectric particles as low-loss alternatives to plasmonic
resonators\cite{ref3}. Nanoholes and aperture arrays emerged as a key
plasmonic platform following the discovery of extraordinary optical
transmission, in which resonant surface modes produce transmission far
exceeding that expected from an isolated subwavelength aperture\cite{ref4}.
Metasurfaces consist of planar arrays of subwavelength meta-atoms whose
geometries control the local amplitude, phase, and polarization of
scattered light. The demonstration of phase discontinuities and
generalized laws of reflection and refraction established a foundation
for flat optical components\cite{ref5}. Photonic crystals use periodic
dielectric modulation to create photonic band gaps that control light
propagation and spontaneous emission, while deliberately introduced
defects can confine light in wavelength-scale, high-\(Q\)
nanocavities\cite{ref6,ref7}. Optical multilayer thin films employ
multiple-beam interference in periodically or resonantly stratified
media. Early theories of wave propagation in periodic structures and
Fabry--Pérot interference provided the physical basis for modern
distributed Bragg reflectors, spectral filters, absorbers, and optical
cavities\cite{ref8,ref9}. Finally, integrated nanophotonic devices confine and
route light through patterned waveguides on a common substrate, evolving
from early integrated optics toward high-index-contrast silicon
platforms supporting compact power splitters, wavelength demultiplexers,
resonators, and computational spectrometers\cite{ref10,ref11}. Together, these
systems span representations ranging from a few continuous
parameters---such as particle radii or layer thicknesses---to
high-dimensional pixelated and freeform geometries, creating
increasingly complex modeling and design problems.

Nanophotonics involves interactions between photons and nanoscale
materials and structures, including coupling to plasmons, which are
collective excitations of an electron or hole charge carrier gas.
Related light--matter and quasiparticle excitations include polaritons,
formed by coupling photons to plasmons or phonons, and polarons, formed
when charge carriers are dressed by local lattice deformations.
Importantly, polaronic interactions have been proposed as a mechanism
for enhancing the critical temperature (\(T_{c}\)) of
superconductors\cite{ref12}. The conventional Bardeen--Cooper--Schrieffer or
Migdal-Eliashberg superconducting state consists of Cooper pairs of
electrons, bound remotely by an effective attraction mediated by
electron-phonon coupling, which is assumed to be weak. A crossover
toward a Bose--Einstein condensate (BEC) can occur as this coupling
increases\cite{ref13}. In a bipolaronic description, the resulting BEC state
may also be viewed as a condensate of bipolarons\cite{ref14}, since each
electron of a Cooper pair, when dressed by a local lattice deformation,
forms a polaron. As the coupling grows, Cooper pairs can carry an
increasingly heavy lattice-distortion cloud, ultimately leading to the
formation of a pinned charge-density wave, an outcome detrimental to
superconductivity. However, low-effective-mass bipolarons can condense
into a bosonic superfluid at moderate coupling, resulting in a
\(T_{c\ }\) exceeding the predictions of Migdal-Eliashberg theory by a
significant margin\cite{ref12}. Recently, some of us proposed a resonant
anti-shielding (RAS) mechanism to maximize this polaronic
enhancement\cite{ref15,ref16}. Because the mechanism depends on several
material and structural parameters, it may also provide a suitable
problem for future AI- and ML-based modeling and optimization. The
generic nanostructure was assumed to be a superlattice of
superconducting monolayers and highly polarizable dielectric films
capable of hosting polaronic modes. The theoretical upper limit of
\(T_{c}\) was modeled using a refined Migdal--Eliashberg framework, in
which the renormalized Eliashberg function is expressed as
\begin{equation}
\alpha^{2}F_{ren}(\omega) = \frac{\alpha^{2}F(\omega)}{\mid \varepsilon_{eff}(\omega) \mid^{2}}
\label{eq:eliashberg}
\end{equation}
where \(\varepsilon_{eff}(\omega)\) is the effective dielectric function
of the substrate. Within this framework, a vanishing
\(\varepsilon_{eff}(\omega)\) indicates the emergence of collective
polarization modes that strongly amplify the superconducting state.

Nanophotonic research can be organized into several closely related
modeling, design, and experimental tasks. Forward modeling starts from a
specified geometry, material distribution, and illumination condition
and solves Maxwell's equations to calculate optical quantities such as
reflection, transmission, scattering, phase, and near- or far-field
distributions\cite{ref17}. Inverse design reverses this direction by
beginning with a desired optical response and searching for one or more
structures capable of producing it\cite{ref18,ref19}. Because distinct
geometries can exhibit similar optical responses, the inverse problem is
generally nonunique, and candidate designs must ultimately be assessed
using EM simulation or experiment\cite{ref19}. Optimization is one approach
to solving this problem, in which geometric parameters or material
distributions are iteratively updated to improve a defined objective;
for a scalar objective, adjoint methods can obtain gradients with
respect to all design variables from one forward and one adjoint solve,
so the number of full-field simulations does not scale with the number
of parameters\cite{ref20}. Fabrication-aware design further incorporates
restrictions such as minimum feature sizes, curvature, permissible
materials, layer thicknesses, and process tolerances so that optimized
structures remain manufacturable and robust\cite{ref21}. Characterization
addresses the complementary inference problem by reconstructing
geometry, refractive index, defects, or complex optical fields from
measured spectra, diffraction patterns, or microscopy data\cite{ref22}. At a
broader level, scientific discovery seeks transferable physical
principles, unexpected structure--property relationships, or testable
hypotheses that extend beyond the optimization of an already specified
optical function\cite{ref23}. Together, these tasks connect structures,
materials, fields, spectra, fabrication, and measurements, providing the
physical problem setting for the AI approaches discussed in the
following sections.

\section{Classical Machine Learning and Deep Learning Foundations}

Machine learning (ML) and deep learning (DL) provide a data-driven
framework for nanophotonic modeling and design by learning
structure--property relationships that are otherwise obtained through EM
simulations or experimental measurements\cite{ref24}. In typical
nanophotonic workflows, ML models may be used in either the forward or
inverse direction. In forward modeling, the input may be a material
composition\cite{ref25}, geometric parameter set\cite{ref24}, topology
image\cite{ref26}, or fabrication condition\cite{ref27}, and the output may be
an optical spectrum, near-field distribution, far-field response, or
figure of merit and so on. In inverse-design settings, the desired
optical response is specified as the input, and the model predicts
candidate structures that satisfy the target functionality\cite{ref28}.
Conventional numerical methods, such as the finite-difference
time-domain (FDTD) method, finite-element method (FEM), and rigorous
coupled-wave analysis (RCWA), compute discretized solutions of Maxwell's
equations for each new structure, whereas trained ML models learn
surrogate mappings from simulation or experimental data and can provide
faster inference after training\cite{ref17,ref29}. Early demonstrations showed
that artificial neural networks (ANNs) can serve as data-driven models
for multilayer nanoparticles\cite{ref24} and plasmonic
nanostructures\cite{ref30}, enabling rapid prediction of scattering or
reflection spectra and supporting inverse retrieval of geometric
parameters. Related deep-learning frameworks were soon extended to
chiral metamaterials\cite{ref31}, tandem inverse-design networks\cite{ref28},
and integrated nanophotonic devices\cite{ref32}, showing that neural
networks can learn nonlinear mappings between optical responses and
increasingly complex photonic structures. More recent progress in
generative modeling\cite{ref33}, attention-based architectures\cite{ref34}, and
large pretrained models\cite{ref35} motivates a broader view of nanophotonic
AI, in which models evolve from task-specific regressors toward
transferable tools for modeling, inverse design, and scientific
discovery\cite{ref36}.

ML methods are commonly divided into supervised and unsupervised
learning. In supervised learning, a model is trained from paired
input--output examples to predict the target variable for new inputs,
whereas unsupervised learning seeks patterns, clusters, or compact
representations from input data without explicit target labels\cite{ref37}.
Classical ML was largely built on models that operate on hand-defined
features, meaning variables chosen by the researcher before training.
Linear regression is the simplest example: it assumes the output can be
approximated as a weighted sum of the input features, where the weights
are learned from data.
\begin{equation}
y\  \approx \mathbf{w}^{T}\mathbf{x} + b\ 
\label{eq:linear}
\end{equation}
Regularization modifies this fitting process by penalizing overly large
or unnecessary weights; ridge regression (also known as
L\textsubscript{2} regularization) shrinks weights smoothly\cite{ref38},
while lasso (L\textsubscript{1} regularization) can force some weights
to zero and therefore performs feature selection\cite{ref39}. Kernel methods
compare data points through nonlinear similarity functions rather than
raw coordinates\cite{ref40}; support vector machines use this idea to find a
decision boundary with the largest separation margin\cite{ref41,ref42}, while
Gaussian process regression treats the unknown function
probabilistically and provides both a prediction and an uncertainty
estimate. Other classical models follow different intuitions: k-nearest
neighbors predicts a new sample from the labels or values of its closest
training examples\cite{ref43}; decision trees make a sequence of
threshold-based decisions\cite{ref44}; and random forests combine many such
decision trees and average their outputs to reduce sensitivity to
noise\cite{ref45}. These methods are often data-efficient and interpretable,
but they depend strongly on the chosen features and become limited when
useful representations must be learned directly from complex data such
as images, sequences, or fields\cite{ref46}.

DL extends classical ML by enabling models to learn internal
representations directly from data, reducing the need for predefined
features. A multilayer perceptron (MLP), or fully connected neural
network (NN), is the simplest DL model. In the (\(l\))-th layer, the
input is the output from the previous layer, (\(\mathbf{x}^{(l - 1)}\)).
This input is multiplied by a learnable weight matrix,
(\(\mathbf{w}^{(l)}\)), shifted by a learnable bias vector,
(\(\mathbf{b}^{(l)}\)), and then passed through a nonlinear activation
function, (\(\sigma\)), to produce the next representation\cite{ref47,ref48},
\begin{equation}
\mathbf{x}^{(l)}\  = \ \sigma\ \left( \mathbf{w}^{(l)}\mathbf{x}^{(l - 1)} + \mathbf{b}^{(l)} \right)
\label{eq:mlp}
\end{equation}
The nonlinear activation is essential because, without it, multiple
linear layers would collapse into a single linear transformation. A
commonly used activation is the rectified linear unit (ReLU), which
keeps positive responses and suppresses negative ones, helping deep
networks train efficiently\cite{ref49}. Because MLPs accept vector inputs,
they are well suited to data-driven modeling when a photonic structure
is described by a small set of geometric or material
parameters\cite{ref24,ref30}. Convolutional neural networks (CNNs) instead use
small trainable filters that slide across image-like inputs, making them
suitable for spatial patterns such as pixelated topologies,
photonic-crystal layouts, microscopy images, and field maps\cite{ref50,ref51}.
For dense image-to-image prediction, encoder--decoder CNNs first
compress the input into a lower-resolution bottleneck and then upsample
it back to the desired output size\cite{ref52,ref53}; U-Net adds skip
connections that pass fine spatial details from the encoder to the
decoder\cite{ref54}. For spectra or time-dependent data, recurrent neural
networks (RNNs) process a sequence step by step while carrying a hidden
state from earlier steps\cite{ref55}; Long short-term memory (LSTM) networks
add gates that control what information is stored, forgotten, or passed
forward, improving learning over long sequences\cite{ref56}. These
architectures became central to photonic data-driven modeling because,
after training, they can approximate nonlinear optical responses much
faster than repeated full-wave simulations\cite{ref24,ref51}.

Generative models are important for inverse design because one target
optical response may correspond to many physically different structures,
making the problem one-to-many rather than a simple regression
task\cite{ref26,ref28}. In this setting, a useful design workflow contains a
generator that proposes candidate structures, an evaluator that
estimates whether their optical responses match the target, and a
criterion that decides how the next candidates should be
improved\cite{ref57,ref58}. Autoencoders first learn to copy their input
through a bottleneck: an encoder compresses the data into a smaller
latent representation, and a decoder reconstructs the original data from
this compressed code\cite{ref59}. In nanophotonics, this latent space
summarizes spectra, fields, or structures for visualization, denoising,
and design-space reduction\cite{ref60}. Variational Autoencoders (VAEs)
modify autoencoders by making the latent code probabilistic, so new
samples can be produced by drawing from the learned latent distribution
and decoding the result\cite{ref61}. This makes VAEs useful for sampling and
optimizing multiple candidate photonic designs rather than searching the
original high-dimensional structure space directly\cite{ref62}. Generative
Adversarial Networks (GANs) use a different idea: a generator creates
synthetic samples, while a discriminator learns to distinguish generated
samples from real examples, and the competition improves the
generator\cite{ref63,ref64,ref65}. In nanophotonic inverse design, GANs generate
metasurface or metagrating geometries that are then checked by a
simulator or neural evaluator\cite{ref26,ref58}. Conditional GANs (cGANs) add
an explicit condition to both generation and discrimination, so the
output is not just realistic but also guided by a target such as a
spectrum, phase profile, or functionality\cite{ref66}. This conditioning is
important for inverse design because the desired optical response must
control the generated structure\cite{ref67}. Wasserstein GANs (WGANs) keep
the generator--discriminator framework but replace the original
adversarial objective with a distance-based criterion, which improves
training stability and reduces mode collapse, where only a narrow subset
of possible designs is produced\cite{ref68}. Diffusion models generate
samples by learning to reverse a gradual noising process, turning random
noise into structured outputs step by step\cite{ref69,ref70}. For
nanophotonics, this denoising-based generation is attractive for diverse
freeform design and for augmenting scarce simulated or experimental
datasets\cite{ref33,ref71}.

Transformers mark another transition in ML because they replace local or
sequential processing with attention, a mechanism that lets each token
compare itself with all other tokens and learn which long-range
relationships are most important\cite{ref34,ref72}. A token can represent many
kinds of data, including a word, an image patch, a spectral segment, a
material label, or a material--thickness layer in a multilayer optical
stack\cite{ref34,ref73,ref74,ref75}. Through large-scale pretraining, transformers can
learn general representations before being adapted to a specific task;
this idea underlies foundation models, which are first trained on broad
datasets and then transferred or fine-tuned for downstream
applications\cite{ref76,ref77}. Unlike conventional supervised models that are
trained for one input--output mapping, foundation models often use
self-supervised or contrastive objectives, such as predicting masked
content or aligning paired data types in a shared latent space\cite{ref78}.
In photonics, this shift has begun with transformer-based models for
optical multilayer inverse design, where a structure can be treated as a
sequence of material and thickness tokens conditioned on target
spectra\cite{ref33,ref36}. More directly for nanophotonics, contrastive
geometry--spectrum pretraining has been proposed to align metasurface
layouts and optical responses within a shared representation
space\cite{ref35}. These examples suggest that the most promising foundation
model for nanophotonics may not be a pure language model, but a
multimodal geometry--spectrum--field--material model trained on
simulations, experiments, and literature, enabling forward prediction,
inverse design, retrieval, transfer learning, and scientific
discovery\cite{ref35,ref76}.

\section{AI for Nanophotonic Forward Modeling}

\begin{figure}[H]
\centering
\includegraphics[width=0.88\linewidth]{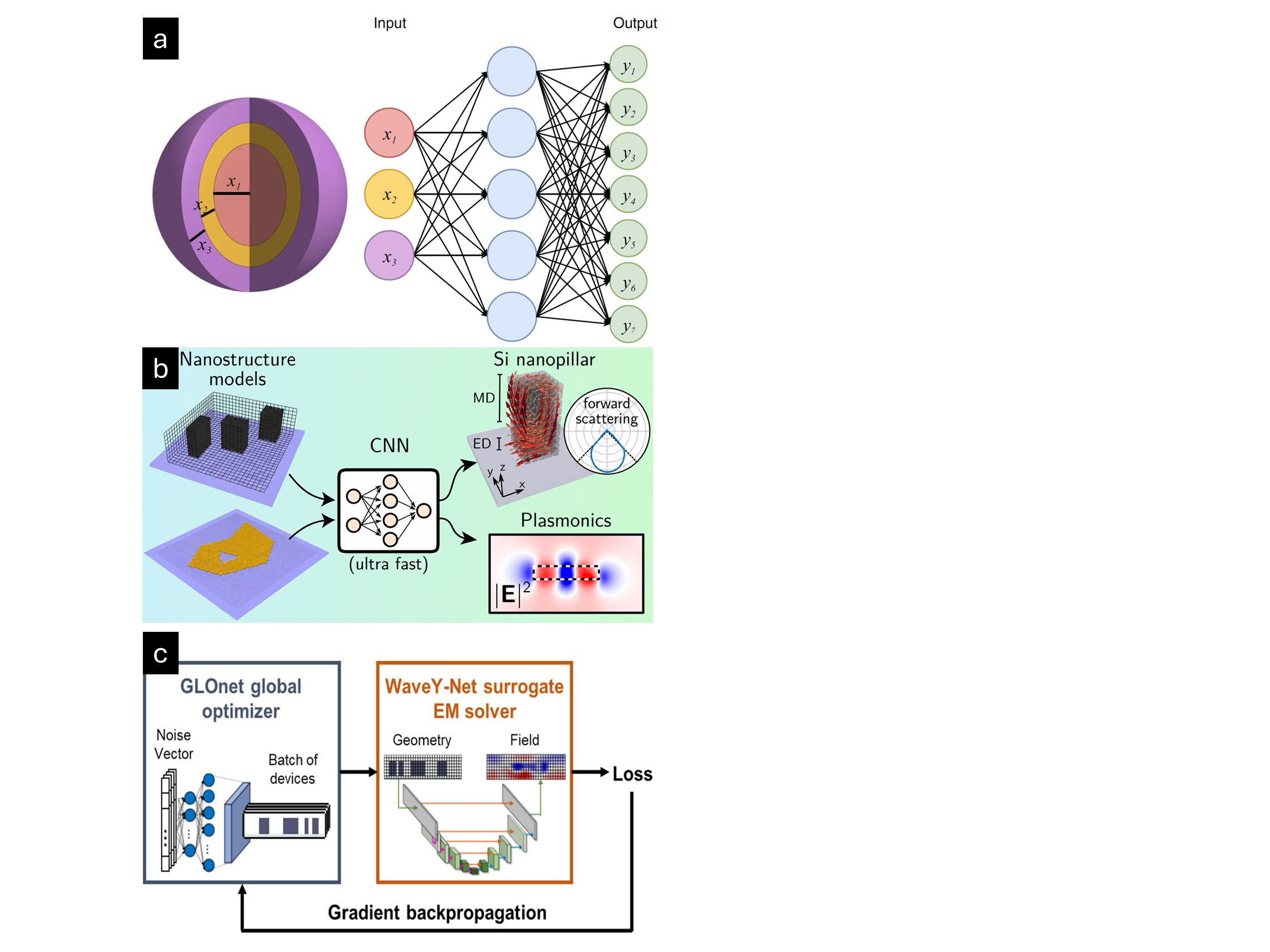}
\caption{AI-based forward modeling in nanophotonics. (a) A parameter-vector model maps nanoparticle geometry variables \(x_{1},x_{2},x_{3}\) to a discretized optical spectrum \(y_{1},y_{2},\ldots,y_{7}\). (b) A convolutional neural network processes spatial nanostructure models to predict near-field and far-field optical responses. (c) A physics-augmented differentiable forward model maps device geometries to electromagnetic fields, allowing the loss to be backpropagated for design optimization. Reproduced with permission from: (a) ref. \cite{ref24}, The American Association for the Advancement of Science; (b) ref. \cite{ref51}, (c) ref. \cite{ref85}, American Chemical Society.}
\label{fig:forward}
\end{figure}

Forward modeling predicts the optical response of a specified structure,
material system, and excitation condition. In nanophotonics, this
mapping is normally obtained from FDTD, FEM, RCWA, or related solvers,
but repeated full-wave calculations become expensive during parameter
sweeps, optimization, and dataset generation. Early AI work therefore
focused on data-driven forward models for low-dimensional, parameterized
structures, where a fully connected NN learns a map from geometric or
material parameters to sampled spectra. As illustrated in Figure~\ref{fig:forward}(a),
Peurifoy, \emph{et al.} used an ANN to map the geometric parameters of
multilayer nanoparticles to wavelength-resolved scattering spectra,
establishing a simple parameter-vector-to-spectrum forward
model\cite{ref24}. Malkiel, \emph{et al.} extended the paradigm to H-shaped
plasmonic nanostructures, combining spectrum prediction with geometry
retrieval from far-field transmission spectra under two
polarizations\cite{ref30}. Ma, \emph{et al}. applied a bidirectional DL
model to 3D chiral metamaterials, predicting reflection and
circular-dichroism spectra from five geometric parameters and revealing
nonlinear links between split-ring geometry and chiroptical
response\cite{ref60}. Liu, \emph{et al}. made the forward model's role more
explicit by embedding a pretrained forward network after an inverse
network, so the forward model acts as a physics evaluator for nonunique
inverse-scattering problems\cite{ref28}. An, \emph{et al}. further advanced
parameterized forward prediction for 3D all-dielectric meta-atoms by
predicting complex transmission components, enabling accurate recovery
of both amplitude and phase\cite{ref50}. Together, these studies established
parameter-vector-to-spectrum prediction as one of the earliest and most
mature applications of AI in nanophotonic forward modeling.

When the input structure is no longer described by a few scalar
parameters, CNN-based models become more natural because they learn
local spatial features directly from images, grids, or voxelized
layouts. Asano and Noda used a neural network with a convolutional layer
to learn the relation between air-hole displacement patterns in 2D
photonic-crystal nanocavities and their Q factors, enabling rapid
gradient-based optimization in a high-dimensional structural
space\cite{ref79}. Sajedian, \emph{et al.} made this image-based view more
explicit by converting plasmonic structures into 2D images; their CNN
extracted spatial features such as shape and position, while an RNN
related these features to the absorption spectrum\cite{ref80}. As
illustrated in Figure~\ref{fig:forward}(b), Wiecha and Muskens used a 3D fully
convolutional encoder--decoder network to map discretized plasmonic and
dielectric nanostructures to internal electric-field distributions, from
which dipole responses, forward-scattering patterns, near-field
intensity, and other optical quantities can be derived\cite{ref51}. An,
\emph{et al.} further applied CNN-based modeling to quasi-freeform
dielectric meta-atoms, combining a 64 $\times$ 64 pattern image with material
index, thickness, and lattice size to predict complex transmission
spectra, amplitude, and phase over a broad frequency range\cite{ref81}.
These studies show that CNNs are not only image classifiers; in
nanophotonics, they can serve as fast optical-response evaluators for
spatially complex structures\cite{ref82}.

Although data-driven forward models can be extremely fast after
training, their predictions are still learned interpolations over the
training distribution. Physics-aware models address this limitation by
incorporating Maxwell's equations, boundary conditions, or field
consistency into the learning process. Early physics-informed neural
network (PINN) studies by Fang and Zhan used residuals of the
frequency-domain Maxwell/Helmholtz equation to solve metamaterial
problems such as cloaking and wave rotation, illustrating how neural
networks can be trained through governing-equation constraints rather
than only labeled simulation pairs\cite{ref83}. MaxwellNet made this idea
more explicit for photonics by using the residual of Maxwell's equations
as a physics-driven loss, training a network to predict electric-field
distributions from material-property maps without requiring full
simulated field labels for every example\cite{ref84}. In WaveY-Net, shown in
Figure~\ref{fig:forward}(c), a U-Net-like model maps dielectric layouts to
magnetic-field distributions, while a discrete Maxwell relation converts
these predictions into electric fields and a Maxwell-consistency loss
penalizes physically inconsistent field patterns\cite{ref85}.
Neural-operator approaches further shift the target from learning one
finite-dimensional mapping to learning families of PDE solution
operators. NeurOLight used a physics-agnostic neural operator to predict
optical fields in parametric photonic devices over varying domains,
wavelengths, sources, and permittivity distributions\cite{ref86}. Similarly,
Fourier neural operators have been used as data-driven EM scattering
solvers with improved data efficiency over U-Net baselines\cite{ref87}. Fast
AI forward models must be judged not only by inference speed but also by
data-generation cost, physical consistency, interpolation range, and
validation against full-wave solvers or experiments; reliable forward
models then become the foundation for inverse design, generative design,
and active learning.

\section{AI for Inverse Design, Optimization, and Generative Design}

Inverse design reverses the forward problem by inferring a structure
from a desired spectrum, phase, or other optical response. Direct
inverse networks learn this response-to-structure mapping and can return
a candidate design in a single inference. Malkiel, \emph{et al.} used
polarization-dependent transmission spectra to predict the geometry of
H-shaped plasmonic nanostructures, with the recovered structure
subsequently evaluated by a spectrum-prediction network\cite{ref30}.
Ghorbani, \emph{et al.} later mapped specified resonance number,
frequency, depth, and bandwidth directly to pixelated metasurface unit
cells operating from 4 to 45 GHz, including a restricted representation
assembled from eight predefined ring motifs\cite{ref88}. For multilayer
photonics, Lininger, \emph{et al.} trained CNNs to infer both material
identities and layer thicknesses from reflectance/transmittance or
ellipsometric spectra, exploring design spaces containing up to
\(10^{12\ }\)possible parameter combinations, as shown in Figure
2(a)\cite{ref89}. A fundamental difficulty, however, is that inverse
scattering is generally nonunique: different structures may produce
nearly identical optical responses, causing conflicting
response-to-structure labels and making direct inverse networks
difficult to train\cite{ref28}. This challenge remains relevant in recent
multichannel systems; a 2025 spin-multiplexed metasurface platform used
a bidirectional network to retrieve silicon-pillar dimensions from
polarization-dependent complex transmission targets while checking the
resulting response with a forward predictor\cite{ref90}. Direct inverse
prediction is therefore extremely fast, but its reliability depends on
how nonuniqueness and structural constraints are handled.

\begin{figure}[H]
\centering
\includegraphics[width=0.88\linewidth]{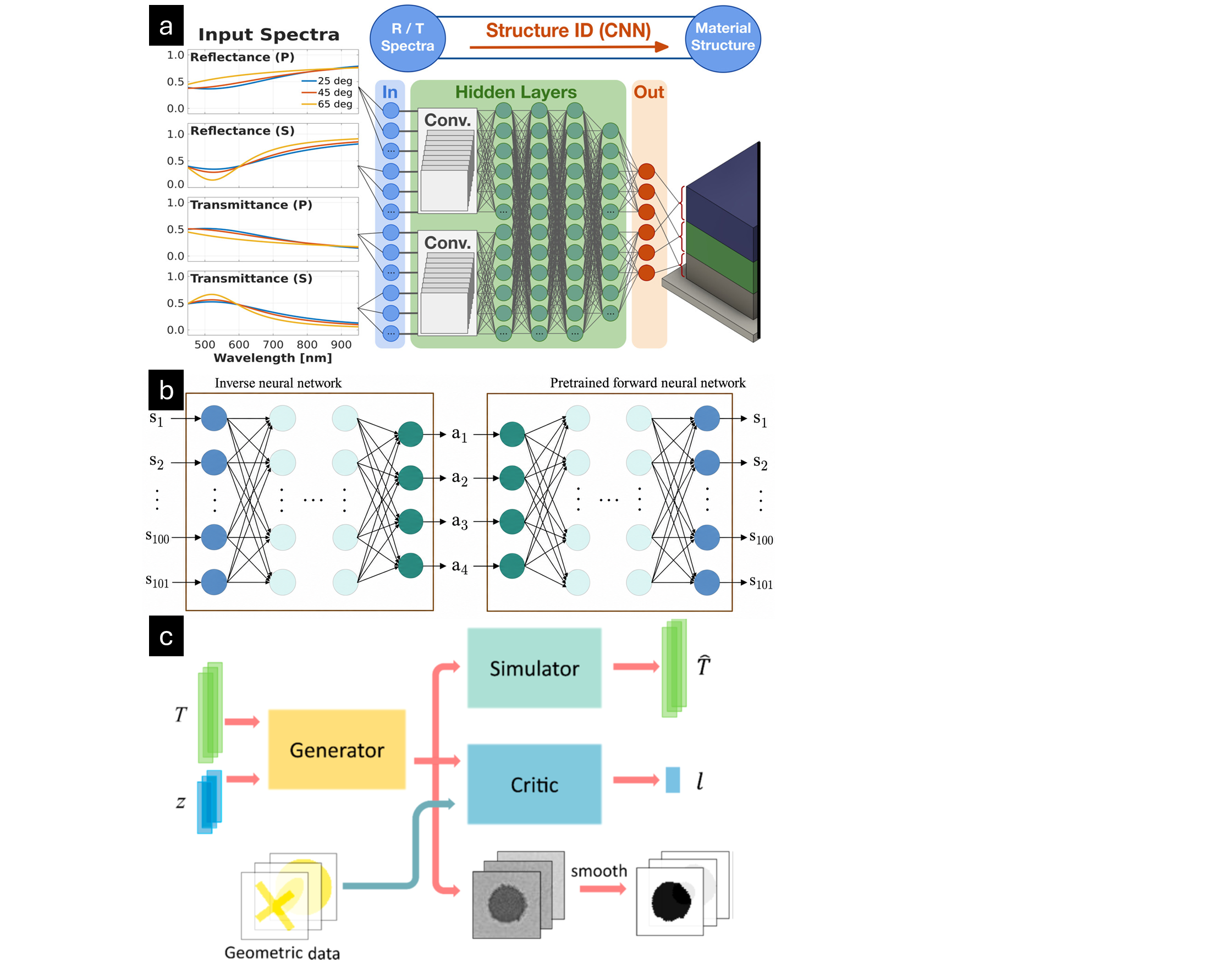}
\caption{Representative AI architectures for inverse nanophotonic design. (a) A convolutional neural network uses wavelength-dependent reflectance and transmittance as inputs; convolutional (``Conv.'') blocks extract spectral features and predict the material sequence of a layered thin film. (b) A tandem network maps the sampled target spectrum \(s_{1},\ldots,s_{101}\)to structural parameters \(a_{1},\ldots,a_{4}\), which are passed through a pretrained forward network to reconstruct \(s_{1},\ldots,s_{101}\). (c) A generative model combines target spectra \(T\) and noise \(z\ \)to produce candidate patterns; the simulator predicts the optical response \(\widehat{T}\), the critic outputs the geometric loss \(l\), and the accepted patterns are smoothed into binary structures. Reproduced with permission from: (a) ref. \cite{ref89}, (c) ref. \cite{ref26}, American Chemical Society; (b) ref. \cite{ref91}, \textcopyright{} Optical Society of America.}
\label{fig:inverse}
\end{figure}

Tandem networks address the nonuniqueness of inverse design by
evaluating a proposed structure in response space rather than requiring
it to reproduce one labeled geometry. In the tandem architecture\cite{ref91}
shown in Figure~\ref{fig:inverse}(b), a target response is first mapped by an inverse
network to candidate structural parameters, which are then passed
through a pretrained forward network; the difference between the
reconstructed and target responses provides the training loss. Liu,
\emph{et al.} introduced this arrangement for nanophotonic structures,
freezing the forward network while optimizing the inverse network so
that any design producing the desired response can be accepted\cite{ref28}.
Xu, \emph{et al}. later applied the same strategy to metasurfaces,
mapping 101 sampled \(S\)-parameter values to four geometric variables
and generating designs within milliseconds\cite{ref91}. Yuan, \emph{et al}.
extended conventional tandem training through a pretrained-combined
network in which the forward and inverse models are jointly fine-tuned,
reducing the domain shift between binary training patterns and the
continuous outputs of the inverse model; the method remained robust with
substantially reduced training data and was experimentally validated in
a compact wavelength demultiplexer\cite{ref92}. Noureen, \emph{et al.}
further incorporated spectral regime, material type, and
fabrication-related geometric limits into a tandem model for UV--visible
meta-atoms, improving data efficiency while predicting both material and
geometry\cite{ref93}. Tandem networks therefore mitigate one-to-many
ambiguity, although their reliability remains limited by forward-model
accuracy, training-domain coverage, and the chosen structural
parameterization.

Inverse design can also retain an explicit optimization loop, with AI
used to evaluate, initialize, or guide candidate structures rather than
predict a final design in one step. Peurifoy, \emph{et al.} optimized
the geometric inputs of a differentiable forward model through
backpropagation, replacing repeated full-wave calculations during the
search\cite{ref24}. GLOnet extended this principle to global topology
optimization by mapping random noise and operating conditions to batches
of metagratings; forward and adjoint simulations supplied efficiency
gradients that shifted the generated device distribution toward
high-performance regions without requiring a precomputed training
set\cite{ref58}. DeepAdjoint adopted a sequential hybrid strategy in which a
pretrained network first generates a promising metasurface, full-wave
simulation validates its response, and adjoint optimization subsequently
refines the geometry beyond the model's training limitations\cite{ref94}.
Bayesian optimization provides a gradient-free alternative: a Gaussian
process estimates both device performance and uncertainty, while an
acquisition function balances exploration of uncertain regions against
exploitation of promising candidates. This strategy produced directional
couplers, Y-splitters, and ring resonators using approximately 100 FDTD
simulations\cite{ref95}. Reinforcement learning instead treats structural
modification as sequential decision-making; a deep Q-network learned
which Si or air cell to flip according to rewards obtained from
RCWA-computed deflection efficiency, enabling exploration of a freeform
metagrating space containing approximately \(10^{17}\) structures
without a prior dataset\cite{ref96}. These methods trade single-pass
inference for iterative physical evaluation, but offer greater
flexibility for complex objectives, fabrication constraints, and
structures outside a fixed response-to-geometry mapping.

Generative models address the one-to-many nature of inverse design by
learning a distribution of possible structures rather than returning
only one deterministic geometry. Figure~\ref{fig:inverse}(c) shows the GAN-based
framework introduced by Liu, \emph{et al}., in which a generator
combines target transmission spectra with a random noise vector to
produce pixelated metasurface patterns, while a pretrained optical
predictor and a critic enforce spectral accuracy and geometrical
plausibility, respectively\cite{ref26}. Ma, \emph{et al}. represented
metamaterial geometries and reflection spectra within a probabilistic
latent space; sampling different latent variables for the same target
response produced multiple topologically distinct designs, while
semi-supervised training incorporated unlabeled geometries to reduce the
required number of EM simulations\cite{ref62}. Tang, \emph{et al}. extended
conditional latent-space generation to integrated nanophotonic power
splitters using an adversarial cVAE, where the desired transmission and
reflection spectra condition the decoder and adversarial censoring
prevents performance information from becoming entangled with the latent
geometry representation\cite{ref97}. Active learning further improved the
model by simulating generated devices and adding the resulting
structure--response pairs back into the training set, enabling compact
power splitters with arbitrary splitting ratios and broadband responses.
More recently, Tanriover, \emph{et al}. learned a compact latent
representation of free-form dielectric meta-atoms generated under
minimum-feature-size and curvature constraints; a genetic algorithm then
searched this latent space to obtain manufacturable polarization filters
and quarter-wave plates\cite{ref98}. Diffusion models provide a newer
alternative by gradually transforming random noise into a structure
conditioned on the required optical response. Hen, \emph{et al}. used
this approach to generate both the binary geometry and height of
diffractive meta-atoms from a target far-field power distribution and
operating wavelength; RCWA gradients could guide the denoising
trajectory for greater physical consistency, or the generated structure
could initialize conventional optimization. Generative inverse design
therefore provides structural diversity and increased design freedom,
but generated candidates must still be checked for optical accuracy,
fabrication feasibility, and generalization beyond the training
distribution.

Despite rapid progress, AI-based inverse design remains strongly
dependent on the structural representation, parameter ranges, and
physical conditions included in the training data; extending a model
across broader shape, thickness, material, polarization, and spectral
spaces while retaining acceptable generalization remains
difficult\cite{ref98,ref99}. Targets outside the learned distribution are
particularly prone to error: the diffusion-based MetaGen model, for
example, showed reduced effectiveness for idealized or
out-of-distribution scattering patterns unless RCWA guidance was
introduced during generation\cite{ref71}. Tandem architectures introduce an
additional source of error because a fixed forward model trained on
simulated binary structures may encounter a domain shift when it
receives the continuous or near-binary outputs of the inverse network,
potentially causing unstable training and inaccurate optical
responses\cite{ref92}. Generative models recover design diversity through
stochastic sampling, but diversity does not guarantee optical fidelity;
sampled latent variables may produce inaccurate structures and therefore
require forward-model screening or full-wave verification\cite{ref62}.
Fabrication feasibility is also essential because unrestricted
geometries can contain gaps, curvatures, or feature sizes that cannot be
reliably manufactured; these constraints can be incorporated into the
geometric representation, training data, or optimization
process\cite{ref21,ref98}. Active learning, simulator-guided generation, and
adjoint refinement improve data efficiency and physical accuracy, but
they reintroduce EM simulations and do not remove the need for numerical
or experimental validation\cite{ref94}. Because most models remain
specialized to particular device families or operating regimes, transfer
learning provides a natural route for migrating learned representations
to related physical scenarios using substantially fewer target-domain
samples\cite{ref100}.

\section{Transfer Learning, Few-Shot Learning, and Data Scarcity}

Transfer learning addresses data scarcity by reusing parameters learned
from a data-rich source problem and fine-tuning them with a smaller
target dataset, rather than training each nanophotonic model from
scratch\cite{ref77,ref101}. Qu, \emph{et al.} first demonstrated that
transferable representations can persist across both closely related and
substantially different optical systems: transferring between multilayer
films with different numbers of layers reduced spectral errors by up to
50.5\%, while selected network layers transferred from nanoparticle
scattering to thin-film transmission still reduced the error by
19.7\%\cite{ref100}. Qiu, \emph{et al}. extended this idea to material-rich
design spaces by combining imbalanced multi-scenario training with
knowledge transfer between multilayer nanoparticles and nanofilms,
obtaining target-domain performance comparable to direct training with
two to three times more samples\cite{ref102}. Fan, \emph{et al.} migrated an
inverse-design network from small-scale to large-scale metasurfaces,
achieving approximately 97\% correlation with target far fields while
reducing the required target data by more than 30\%\cite{ref103}. Transfer
is particularly effective when the source and target share both physical
mechanisms and compatible structural representations. For example, an
ANN trained on 5040 annular aperture array samples was fine-tuned using
only 525 nanohole array samples by treating the nanohole array as a
special case of the annular aperture array with an inner diameter of
zero; the transferred model converged in 35 rather than 146 epochs and
reduced the validation loss by 37\% relative to training from
scratch\cite{ref104,ref105}. Beyond accelerating modeling and design, these
results indicate that transfer learning can expose reusable
representations of optical physics, helping identify which
structure--response relationships persist across device families and
thereby supporting scientific discovery.

Transfer learning can also bridge larger changes in task definition,
operating conditions, and material systems. Zhu, \emph{et al.}
transferred visual features learned from ImageNet to metasurface
modeling by treating binary meta-atom patterns as images and fine-tuning
an Inception-V3 network for reflection-phase classification; the
resulting phase library enabled rapid generation of focusing and
anomalous-reflection metasurfaces with approximately 90\% prediction
accuracy\cite{ref106}. Peng, \emph{et al}. examined transfer more
systematically for large metasurface arrays by changing the incident
angle, polarization, element material, and geometry of a pretrained
scattering model. Transfer consistently improved far-field prediction
over random initialization, while the required target data depended
strongly on the physical similarity between the source and target
problems; in the most favorable case, a 3\% error was reached with
roughly three orders of magnitude fewer samples\cite{ref107}. Transfer can
also span widely separated spectral regions. Xu, \emph{et al.}
pretrained a complex-valued network using approximately 60,000 infrared
samples and adapted it to the terahertz range using only 1000 target
samples, improving prediction accuracy by about 26\% compared with
training from scratch\cite{ref108}. More recently, Wang, \emph{et al.} used
the shared Drude description of aluminum, gold, and silver to define
physically motivated material similarity, transferring a model trained
on aluminum metasurfaces to gold and silver systems while reducing the
required target data by 50\%\cite{ref109}. These studies show that
successful transfer depends not only on network architecture, but also
on identifying a meaningful physical connection between the source and
target domains.

Data scarcity can also be addressed without transferring an entire
pretrained model. Data augmentation expands the effective training set
by constructing additional physically valid examples from existing
simulations. NeurOLight exploited the linearity of Maxwell's equations
through superposition-based augmentation, combining single-source field
solutions into multi-source examples and maintaining useful prediction
accuracy even when the available training set was substantially
reduced\cite{ref86}. Self-supervised and semi-supervised methods instead
learn structural information from geometries that lack simulated optical
labels. Ma and Liu dynamically generated unlabeled metasurface patterns
during training and used the model's predicted spectra as internal
supervisory signals for reconstruction; compared with fully supervised
training, this approach reduced the total test loss by 14.4\% and the
spectral prediction error by approximately 15\%, while producing more
clearly separated geometry clusters in the learned latent
space\cite{ref110}. Active learning takes a different approach by deciding
which candidates are most valuable to simulate next. Tang, \emph{et al.}
used a preliminary generative model to produce 1000 new power-splitter
patterns, labeled them using FDTD, and appended them to the training
set, improving the performance of subsequently generated
devices\cite{ref97}. Singh, \emph{et al}. later applied active learning to a
much broader metasurface space containing 36 shape classes together with
variable thickness and pitch. A combination of committee-based
uncertainty sampling and latent-space clustering selected both
informative and geometrically diverse structures for simulation,
achieving accuracy comparable to random sampling with only 42,000
labeled examples instead of 240,000---an 82\% reduction in data
generation\cite{ref99}. These approaches are complementary: augmentation
reuses existing simulations, self-supervision extracts information from
unlabeled structures, and active learning concentrates new simulations
in the most informative regions of the design space.

Related data-scarcity challenges arise in the search for superconducting
heterostructures governed by electron--phonon and polaronic
interactions. Evaluating the renormalized Eliashberg function across
many candidate material pairings requires repeated density functional
theory (DFT) and density functional perturbation theory (DFPT)
calculations, which are difficult to scale. Moreover, generalization to
novel layered heterostructures remains challenging because polaronic,
bipolaronic, and superconducting behavior involve coupled physical
mechanisms that are only sparsely represented in available training
data. Graph-based and Hamiltonian-learning models offer complementary
ways to reduce this computational cost. The Atomistic Line Graph Neural
Network (ALIGNN) represents crystal structures using atomic
connectivity, bond distances, and bond angles and has been applied to a
broad range of materials-property predictions\cite{ref111}. BETE-NET uses an
ensemble of equivariant graph neural networks to predict the Eliashberg
spectral function \(\alpha^{2}F(\omega)\) from crystal structures. By
incorporating the site-projected phonon density of states as a
physics-informed input, it improves learning from limited data and
supports high-throughput screening for electron--phonon
superconductors\cite{ref112}. Machine-learned Hamiltonian approaches provide
another route. HamEPC predicts the electronic Hamiltonian and its
gradients with respect to atomic coordinates using an
\(E(3)\)-equivariant graph neural network, allowing electron--phonon
coupling matrices to be evaluated without repeating the full
self-consistent DFT procedure\cite{ref113}.

Few-shot learning seeks to adapt a model using only a small number of
costly target-domain samples and is therefore especially useful when
high-fidelity simulations or experiments are limited. Gao, \emph{et al}.
encoded inexpensive low-fidelity simulations into the prior of a
Bayesian regression model and then updated it using high-fidelity data.
For silicon waveguides and Y-branches under process variation, useful
performance models were obtained with only ten expensive simulations,
yielding more than sevenfold lower error than conventional regression or
an MLP using the same high-fidelity data\cite{ref114}. Huang, \emph{et al.}
pretrained a recurrent holographic reconstruction network on several
specimen types and adapted it to previously unseen samples using only 80
image fields of view. Freezing the reusable recurrent backbone reduced
the number of trainable parameters by approximately 90\% and accelerated
convergence by about 2.5 times\cite{ref115}. Meta-learning goes further by
training a model specifically to adapt rapidly. Yang, \emph{et al.}
incorporated possible fabrication deviations into simulated
meta-training tasks and calibrated a fabricated computational
microspectrometer using only 15 measured color samples, reducing the
spectral reconstruction error from \(7.2 \times 10^{- 3}\) to
\(1.2 \times 10^{- 3}\)\(\lbrack 116\rbrack\). Shin, \emph{et al}.
similarly transferred knowledge from simulation-trained OLED models to
modified multilayer structures and showed, using synthetic experimental
datasets, that systematic simulation--experiment discrepancies could be
learned with only a few dozen target samples\cite{ref117}. These studies
motivate broader pretrained models whose representations can be reused
across structures, measurements, and physical conditions through
lightweight fine-tuning or few-shot adaptation, providing a natural
transition from transfer learning to foundation models.

\section{Foundation Models and Future Directions for Nanophotonics}

Here, ``foundation model'' denotes a model pretrained on heterogeneous
or large-scale domain data whose representation is demonstrably reusable
across multiple downstream tasks, device families, or operating
conditions through prompting or adaptation, rather than a model trained
for one fixed input--output mapping\cite{ref76}. For a large language model
(LLM), a sentence describing the photoelectric effect is first divided,
according to its tokenizer, into subwords, characters, punctuation
marks, or mathematical symbols, and each token is mapped to a numerical
embedding\cite{ref118}. A transformer then uses self-attention to update
each embedding according to its relationships with the surrounding
tokens, allowing terms such as photon energy, frequency, work function,
electron emission, and \(E = h\nu\) to be represented in
context\cite{ref34}. During autoregressive pretraining, the model learns the
probability distribution of the next token and adjusts its parameters by
minimizing cross-entropy over a large text corpus\cite{ref119,ref120}. Repeated
exposure to scientific explanations can therefore produce internal
representations that encode relationships among physical concepts, and
recent theory suggests that, under specific identifiability assumptions,
next-token representations may approximate linear transformations of the
posterior probabilities of latent human-interpretable concepts\cite{ref119}.
However, encoding a concept is not equivalent to possessing a physically
grounded understanding of it: a text-only model learns relationships
within human-generated descriptions rather than interacting directly
with photons, electrons, instruments, or materials\cite{ref121}. Such a
model may retrieve known laws, connect equations, and propose plausible
hypotheses about the photoelectric effect, but a genuinely new physical
phenomenon must be expressed as a falsifiable prediction and validated
through calculation, simulation, or experiment\cite{ref122}. Thus,
next-token pretraining can compress established scientific knowledge
into reusable representations, whereas reliable scientific discovery
requires connecting those representations to physical models, tools,
observations, and experimental feedback\cite{ref76}.

Current photonic models described as foundation models remain
domain-specific, but they demonstrate how broad pretraining can reduce
the need for separate networks for narrowly defined design targets.
OptoGPT represents each multilayer layer as a material--thickness token
and serializes stacks with different materials and layer numbers into
variable-length sequences\cite{ref36}. Trained on ten million structures
simulated by the transfer-matrix method, its decoder-only transformer
autoregressively generates multilayer stacks from target reflection and
transmission spectra and can be adapted to different incidence
conditions, polarizations, and fabrication constraints through
fine-tuning or constrained token sampling\cite{ref36}. OptoLlama extends
this sequence formulation using masked diffusion: it begins with a fully
masked stack and reconstructs all material--thickness tokens iteratively
while conditioning on reflectance, absorptance, and transmittance
spectra\cite{ref33}. By considering the complete stack during denoising
rather than committing to layers one by one, OptoLlama reduced the mean
spectral error by 3.45-fold relative to a reproduced OptoGPT baseline
and generated multiple physically plausible designs through stochastic
sampling. MOCLIP instead uses contrastive learning to align CNN-encoded
metasurface geometries and MLP-encoded spectra in a shared latent space.
It was pretrained on 466,537 experimentally fabricated and measured
silicon metasurfaces spanning 42 geometric degrees of freedom, enabling
zero-shot geometry--spectrum retrieval and latent-space optimization
without relying solely on simulated data\cite{ref35}. General-purpose LLMs
have also been adapted through prompting and parameter-efficient
fine-tuning to generate simulation code and learn text-encoded mappings
between metasurface parameters and spectra, although such approaches do
not yet constitute broad nanophotonic foundation models\cite{ref123}.
Collectively, under the operational criterion above, these studies are
best viewed as candidate domain foundation models: they reuse pretrained
representations within multilayer or metasurface design spaces, but they
do not yet jointly model the diversity of geometries, materials, EM
fields, fabrication data, experiments, and scientific literature
required for a ChatGPT-level multimodal nanophotonic system.

Recent AI-for-science research suggests that scientific foundation
models will emerge not from text alone, but from combining physical
simulation, large-scale pretraining, generative modeling, active
learning, automated experiments, and agentic reasoning. DFT is an
important foundation for this development in materials science because
it provides first-principles approximations of electronic energies and
forces, although it is an electronic-structure framework rather than a
universal model of all physical phenomena\cite{ref124,ref125}. GNoME embedded
DFT within an active-learning loop: graph networks screened candidate
crystals, selected structures were evaluated by DFT, and the resulting
calculations were returned to the training set, leading to 2.2 million
structures predicted to be stable relative to previous databases and
381,000 entries on the updated convex hull\cite{ref124}. MatterGen moved
from candidate screening to conditional generation by pretraining a
diffusion model on 607,683 stable crystal structures and fine-tuning it
through adapters to generate materials with specified chemistry,
symmetry, and mechanical, electronic, or magnetic properties; one
generated material was experimentally synthesized with a measured
property within 20\% of the target value\cite{ref126}. A-Lab further closed
the loop between computation and experiment by combining ab initio
databases, literature-trained synthesis models, robotic preparation and
characterization, and active learning, successfully synthesizing 36 of
57 targeted inorganic compounds during 17 days of autonomous
operation\cite{ref127,ref128}. Beyond materials science, Aurora was pretrained
on more than one million hours of heterogeneous geophysical data and
subsequently fine-tuned for air-quality, ocean-wave, tropical-cyclone,
and high-resolution weather forecasting, demonstrating that a shared
physical representation can support several related scientific
tasks\cite{ref129}. Co-Scientist extends this progression from prediction
and generation toward scientific reasoning by using multiple specialized
agents to search the literature, generate, critique, rank, and evolve
hypotheses, with selected biomedical proposals subsequently subjected to
experimental validation\cite{ref130,ref131}. Together, these developments
suggest that a future nanophotonic foundation model should not merely be
a larger geometry--spectrum predictor, but a scientist-in-the-loop
platform that connects EM simulations, literature and experimental data,
generative design, uncertainty-guided candidate selection, fabrication,
characterization, and iterative revision of physical hypotheses.

A ChatGPT-level nanophotonic system would require a general multimodal
model that can interpret and generate text, equations, material
descriptions, parameterized and freeform geometries, spectra, EM fields,
fabrication records, and experimental measurements within a unified
interface\cite{ref132}. Each modality would first be converted into a
compatible representation: text and equations into discrete tokens,
multilayer structures into material--thickness sequences, layout and
field images into patch or latent tokens, and spectra into
wavelength-indexed numerical embeddings containing units, polarization,
incidence conditions, and material metadata\cite{ref74,ref118}. Contrastive
training could then align paired modalities---such as geometry and
spectrum, structure and field, simulation and measurement, or figures
and scientific descriptions---whereas masked reconstruction, next-token
prediction, and cross-modal generation would teach the model to infer
missing information and translate between
representations\cite{ref73,ref78,ref133}. The resulting embeddings could be
processed by a central transformer trained on interleaved multimodal
records and connected to modality-specific decoders for generating
explanations, spectra, fields, structures, or executable simulation
instructions\cite{ref123,ref134}. Physical grounding should be introduced
through objectives that penalize violations of Maxwell's equations,
energy conservation, reciprocity, symmetry, boundary conditions, and
causal spectral relationships, while exact calculations are delegated to
trusted numerical solvers rather than approximated solely through token
generation\cite{ref85}. Tool-use training would teach the model when and how
to call FDTD, FEM, RCWA, transfer-matrix, adjoint-optimization, CAD,
material-database, literature-retrieval, and uncertainty-estimation
tools and how to incorporate their verified outputs into subsequent
reasoning\cite{ref135}. Post-training could combine expert instruction
tuning, preference feedback, self-critique, and multi-agent hypothesis
evaluation, progressing from forward modeling and physical explanation
to inverse design, fabrication planning, and falsifiable hypothesis
generation\cite{ref130}. Solver and experimental outcomes would then form a
closed feedback loop in which failed predictions trigger targeted
simulations or measurements and provide evidence for subsequent model
improvement\cite{ref124,ref127}. Such a system should be evaluated through
zero- and few-shot transfer to unseen geometries, materials,
wavelengths, and tasks, together with calibrated uncertainty,
physical-law compliance, reproducible tool traces, and independent
simulation or experimental validation\cite{ref129,ref130}. Thus, a GPT-level
nanophotonic foundation model would be not a single monolithic network,
but an integrated scientific system combining multimodal
representations, physical models, external tools, memory, verification,
and human oversight.

Major challenges remain before such a multimodal system can be trusted
for nanophotonic research, because scientific foundation models must
integrate heterogeneous data without losing physical meaning,
uncertainty information, or transferability\cite{ref136,ref137}. Tokenization
and latent compression must preserve continuous geometries, amplitudes
and phases, complex-valued fields, units, and sharp spectral or spatial
features rather than reducing them to approximate visual patterns;
recent physical-science tokenizers and dual-scale multimodal models
demonstrate the importance of retaining both numerical magnitude and
fine-scale structure\cite{ref138,ref139}. Cross-modal training must also align
simulations, measurements, literature, and fabrication records despite
missing modalities, instrument-specific noise, incompatible resolutions,
and differences in physical conditions\cite{ref140,ref141}. More critically,
fluent outputs can remain incorrect or unsupported, so the system should
estimate its uncertainty, invoke verified numerical or experimental
tools when needed, document its assumptions and execution settings, and
distinguish an unverified model prediction from a solver- or
experiment-validated result\cite{ref142,ref143,ref144,ref145}. Evaluation should separately
test zero- and few-shot transfer under controlled distribution shifts,
compliance with physical and numerical constraints, quantitative
accuracy, uncertainty calibration, and reproducibility of the complete
computational workflow\cite{ref146,ref147,ref148}. Finally, novelty alone should not
be treated as scientific discovery: the system should formulate a
testable hypothesis, design a discriminating simulation or experiment,
interpret the resulting evidence, and withstand independent
validation\cite{ref122,ref149}.

\section{Summary and Conclusions}

Nanophotonic modeling and design have progressed from physics-based
simulations and iterative optimization toward increasingly general
data-driven systems. Early ML and DL methods learned mappings for
specific structures and tasks, enabling rapid forward modeling and
direct or optimization-assisted inverse design. Generative models
subsequently addressed the nonuniqueness of inverse problems by
producing multiple candidate structures, while transfer learning,
few-shot learning, and physics-aware training reduced the dependence on
large task-specific datasets. More recently, transformers and domain
foundation models have begun to unify variable structures, optical
responses, and design objectives within reusable pretrained
representations. However, current models remain specialized compared
with general multimodal AI systems. Progress toward a broadly capable
nanophotonic foundation model will require joint representations of
structures, materials, spectra, EM fields, fabrication conditions,
measurements, equations, and scientific literature, together with
reliable access to numerical solvers and experimental tools. Such
systems should not replace EM theory or physical validation, but rather
connect them through scalable prediction, design generation, uncertainty
estimation, and hypothesis testing. Substantial advances may therefore
come not from a single larger neural network, but from integrated
human--AI workflows in which learned models, physical solvers,
fabrication, characterization, and scientific reasoning continuously
inform one another.

\section*{Data availability statement}

No new data were created or analysed in this study.

\section*{ORCID iDs}

\noindent Chaobin Yang: \href{https://orcid.org/0000-0002-7550-8154}{0000-0002-7550-8154}\\
Xueqing Liu: \href{https://orcid.org/0009-0009-0637-4357}{0009-0009-0637-4357}\\
Yiqun Fu: \href{https://orcid.org/0009-0009-3628-4483}{0009-0009-3628-4483}\\
Fengbo Zhou: \href{https://orcid.org/0000-0002-6852-2994}{0000-0002-6852-2994}\\
Krzysztof Kempa: \href{https://orcid.org/0000-0003-3953-1485}{0000-0003-3953-1485}\\
Stefano~Anzellotti: \href{https://orcid.org/0000-0002-8964-6988}{0000-0002-8964-6988}\\
Michael J. Naughton: \href{https://orcid.org/0000-0002-6733-2398}{0000-0002-6733-2398}

\end{document}